\documentclass{aa}
\usepackage{aalongtable}
\usepackage{graphicx}
\usepackage{txfonts}
\usepackage{natbib}
\usepackage[figuresright]{rotating}
\usepackage[a4paper,breaklinks]{hyperref}

\begin{document}
\def\teff{$T\rm_{eff}$ }
\def\kms {\,$\mathrm{km\, s^{-1}}$ }
\def\kmss {\,$\mathrm{km\, s^{-1}}$}
\def\ms {$\mathrm{m\, s^{-1}}$ }

\newcommand{\Teff}{\ensuremath{T_\mathrm{eff}}}
\newcommand{\g}{\ensuremath{g}}
\newcommand{\gf}{\ensuremath{gf}}
\newcommand{\loggf}{\ensuremath{\log\gf}}
\newcommand{\glog}{\ensuremath{\log\g}}
\newcommand{\pun}[1]{\,#1}
\newcommand{\cobold}{\ensuremath{\mathrm{CO}^5\mathrm{BOLD}}}
\newcommand{\linfor}{Linfor3D}
\newcommand{\xx}{\ensuremath{\mathrm{1D}_{\mathrm{LHD}}}}
\newcommand{\punms}{\mbox{\rm\,m\,s$^{-1}$}}
\newcommand{\punkms}{\mbox{\rm\,km\,s$^{-1}$}}
\newcommand{\abuhe}{\mbox{Y}}
\newcommand{\grav}{\ensuremath{g}}
\newcommand{\mlp}{\ensuremath{\alpha_{\mathrm{MLT}}}}
\newcommand{\mlpcm}{\ensuremath{\alpha_{\mathrm{CMT}}}}
\newcommand{\moh}{\ensuremath{[\mathrm{M/H}]}}
\newcommand{\senv}{\ensuremath{\mathrm{s}_{\mathrm{env}}}}
\newcommand{\shelio}{\ensuremath{\mathrm{s}_{\mathrm{helio}}}}
\newcommand{\smin}{\ensuremath{\mathrm{s}_{\mathrm{min}}}}
\newcommand{\spun}{\ensuremath{\mathrm{s}_0}}
\newcommand{\sstar}{\ensuremath{\mathrm{s}^\ast}}
\newcommand{\tauross}{\ensuremath{\tau_{\mathrm{ross}}}}
\newcommand{\ttaurelation}{\mbox{T$(\tau$)-relation}}
\newcommand{\Ysurf}{\ensuremath{\mathrm{Y}_{\mathrm{surf}}}}
\newcommand{\mD}{\ensuremath{\left\langle\mathrm{3D}\right\rangle}}

\newcommand{\draftflag}{false}

\newcommand{\beq}{\begin{equation}}
\newcommand{\eeq}{\end{equation}}
\newcommand{\pdx}[2]{\frac{\partial #1}{\partial #2}}
\newcommand{\pdf}[2]{\frac{\partial}{\partial #2}\left( #1 \right)}

\newcommand{\var}[1]{\ensuremath{\sigma^2_{#1}}}
\newcommand{\sig}[1]{\ensuremath{\sigma_{#1}}}
\newcommand{\cov}[2]{\ensuremath{\mathrm{Cor}\left[#1,#2\right]}}
\newcommand{\xtmean}[1]{\ensuremath{\left\langle #1\right\rangle}}

\newcommand{\eref}[1]{\mbox{(\ref{#1})}}

\newcommand{\Vact}{\ensuremath{\nabla}}
\newcommand{\Vad}{\ensuremath{\nabla_{\mathrm{ad}}}}
\newcommand{\Veddy}{\ensuremath{\nabla_{\mathrm{e}}}}
\newcommand{\Vrad}{\ensuremath{\nabla_{\mathrm{rad}}}}
\newcommand{\Vraddiff}{\ensuremath{\nabla_{\mathrm{rad,diff}}}}
\newcommand{\cp}{\ensuremath{c_{\mathrm{p}}}}
\newcommand{\taueddy}{\ensuremath{\tau_{\mathrm{e}}}}
\newcommand{\vconv}{\ensuremath{v_{\mathrm{c}}}}
\newcommand{\Fconv}{\ensuremath{F_{\mathrm{c}}}}
\newcommand{\lmix}{\ensuremath{\Lambda}}
\newcommand{\Hp}{\ensuremath{H_{\mathrm{P}}}}
\newcommand{\Hptop}{\ensuremath{H_{\mathrm{P,top}}}}
\newcommand{\COBOLD}{{\sc CO$^5$BOLD}}

\newcommand{\changed}{}

\newcommand{\I}{\ensuremath{I}}
\newcommand{\Irot}{\ensuremath{\tilde{I}}}
\newcommand{\F}{\ensuremath{F}}
\newcommand{\Frot}{\ensuremath{\tilde{F}}}
\newcommand{\vsini}{\ensuremath{V\sin(i)}}
\newcommand{\vvsini}{\ensuremath{V^2\sin^2(i)}}
\newcommand{\vsinimu}{\ensuremath{\tilde{v}}}
\newcommand{\rotint}{\ensuremath{\int^{+\vsinimu}_{-\vsinimu}\!\!d\xi\,}}
\newcommand{\imu}{\ensuremath{m}}
\newcommand{\imupone}{\ensuremath{{m+1}}}
\newcommand{\nmu}{\ensuremath{N_\mu}}
\newcommand{\msum}[1]{\ensuremath{\sum_{#1=1}^{\nmu}}}
\newcommand{\wmu}{\ensuremath{w_\imu}}

\newcommand{\tchar}{\ensuremath{t_\mathrm{c}}}
\newcommand{\Nt}{\ensuremath{N_\mathrm{t}}}

\title{Accuracy of spectroscopy-based radioactive dating of stars}
\subtitle{ }

\author{
H.-G. Ludwig \inst{1,2};
E. Caffau    \inst{2};
M. Steffen   \inst{3};
P. Bonifacio \inst{1,2,4};
L. Sbordone \inst{1,2};
}
\institute{CIFIST Marie Curie Excellence Team
\and
GEPI, Observatoire de Paris, CNRS, Universit\'e Paris Diderot; Place
Jules Janssen 92190
Meudon, France
\and
Astrophysikalisches Institut Potsdam, An der Sternwarte 16, D-14482 Potsdam, Germany
\and
Istituto Nazionale di Astrofisica,
Osservatorio Astronomico di Trieste,  Via Tiepolo 11,
I-34143 Trieste, Italy
}
\authorrunning{Ludwig et al.}
\titlerunning{Accuracy of spectroscopy-based radioactive stellar dating}
\offprints{Hans.Ludwig@obspm.fr}
\date{Received ...; Accepted ...}

\abstract
{Combined spectroscopic abundance analyses of stable and radioactive elements can be
  applied for deriving stellar ages. The achievable precision depends on
  factors related to spectroscopy, nucleosynthesis, and chemical evolution.}
{We quantify the uncertainties arising from the spectroscopic
  analysis, and compare these to the other error sources.} 
{We derive formulae for the age uncertainties arising
  from the spectroscopic abundance analysis, and apply them to spectroscopic
  and nucleosynthetic data compiled from the literature for the Sun and
  metal-poor stars.}
{We obtained ready-to-use analytic formulae of the age uncertainty for the
  cases of stable+unstable and unstable+unstable chronometer pairs, and
  discuss the optimal relation between to-be-measured age and mean lifetime of
  a radioactive species. Application to the literature data indicates that, for
  a single star, the achievable spectroscopic accuracy is limited to about
  $\pm$20\,\% for the foreseeable future. At present, theoretical
  uncertainties in nucleosynthesis and chemical evolution models form the
  precision bottleneck. For stellar clusters, isochrone fitting provides a
  higher accuracy than radioactive dating, but radioactive dating becomes
  competitive when applied to many cluster members simultaneously, reducing
  the statistical errors by a factor $\sqrt{N}$.}
%
{Spectroscopy-based radioactive stellar dating would benefit from improvements
  in the theoretical understanding of nucleosynthesis and chemical evolution.
  Its application to clusters can provide strong constraints for
  nucleosynthetic models.}

\keywords{Sun: abundances -- Stars: abundances -- Stars: evolution -- Stars: fundamental parameters}
\maketitle


\section{Introduction}

The determination of absolute ages for stars is of paramount importance in
astrophysics since ages allow the timescales on which stellar populations and
their chemical composition evolve to be established.  The presence of
long-lived unstable elements in stellar atmospheres offers a way to date stars
by investigating their radioactive decay spectroscopically. In its simplest
form, one determines the photospheric abundance of a pair of elements, of
which one is unstable. By means of theoretical modeling of the
nucleosynthetic production channels for the two elements, the original
abundance of the unstable nucleus can be estimated from the abundance of the
stable one. This allows the time-span over which the unstable
element decayed to be determined.

The idea emerged early after the discovery of the rapid neutron capture
process (short $r$-process) as the source of production of radionuclides such
as thorium and uranium \citep[see][and references therein for an account of
its early applications]{cowan91}. Since then it has been applied in a number
of cases \citep[among the most recent][]{frebel07} but, typically, as a ``side
product'' of analyses of the more general n-capture enrichment patterns.  To
our knowledge, the only effort in recent years to exploit radioactive dating
to systematically derive the age of a significant sample of stars has been the
one by \citet{delpeloso05,delpeloso2,delpeloso3}.

One interest in such dating techniques lies in its applicability to field
stars for which no distance information, hence no accurate information on
surface gravity or absolute luminosity, is available, so that they cannot be
placed with confidence on theoretical isochrones in the Hertzsprung-Russell
diagram (HRD). 
For cluster stellar populations, on the other hand, HRD fitting
usually provides rather precise absolute ages \citep[e.g.][]{vandenberg96}.
However, for stellar clusters (especially globular clusters), radioactive dating
could produce an age estimate of comparable or even higher precision if
applied to a large enough sample of stars, and would as such be very valuable
as a test of the theoretical stellar isochrones.

Although potentially effective, spectroscopy-based radioactive dating is
difficult to apply. The most easily measured long-lived radioactive element,
thorium, is only accessible (with the exception of strongly $r$-process
enriched stars) through a single, weak line in the optical spectrum of stars,
in which high accuracy is needed in the measurement to provide a reliable age.
Besides lying on the red wing of a Fe-Ni blend, this unique \ion{Th}{ii} line
is itself also blended by weaker lines of vanadium and cobalt.  The relative
strength of the various components for the Sun are discussed in
\citet{caffau08hfth} where the line-list of \citet{delpeloso05} was applied.
Although rather easily measured, the most frequently used reference element,
europium, is believed to constitute a less-than-optimal reference. In fact,
its nucleosynthetic ties with Th are rather weak, and the production ratio is
sensitive to the originating supernova characteristics
\citep[][]{kratz07,farouqi08}.

Besides the astrophysical problem that
  the $r$-process site has not been unambiguously identified, it is unclear 
  how universal the resulting abundance pattern of the
  $r$-process nucleosynthesis actually is.  In view of the complex nuclear
  reactions shaping the $r$-process it is not obvious that the outcome is the
  same for a potentially wide range of the physical conditions governing the
  formation site(s) \citep{goriely99,arnould07,sneden08,roederer09}.  To
  minimize the influence of poorly constrained physical conditions at the
  formation site it is desirable to use a reference element with an atomic mass as
  similar as possible to thorium. This should reduce the
  uncertainties in the formation properties to the uncertainties
  related to the properties of the involved nuclei as such, and environmental
  factors should cancel out. As pointed out before by various authors
  \citep[e.g.,][]{goriely99} the Th/U pair is in this sense optimal, however,
  again difficult to apply since uranium is difficult to detect
  spectroscopically.  Recently, \citet{kratz07} suggested the element hafnium
  as suitable reference element for thorium-based radioactive dating. Hafnium
  is closer in mass to thorium than europium, but more importantly, Kratz and
  collaborators showed by detailed modeling that the production ratio Th/Hf
  is not strongly varying over a wide range of neutron fluxes during the
  $r$-process. Since hafnium is a spectroscopically accessible element
  this makes it an interesting candidate as reference element.

Observationally, the forthcoming introduction of Extremely Large
Telescopes (ELTs) is likely to make radioactive dating of stars a much more
exploitable technique, since the high signal-to-noise (S/N) ratio required is
one of the main difficulties associated with such studies. In this
perspective, a precise assessment of the uncertainties associated with such
technique is currently lacking, and attempted here.  We derive the formulae
for the precision to which spectroscopy can provide age estimates. They are
subsequently applied to abundance analyses of stars compiled from the
literature.  This paper is a ``spin-off'' of our work \citep{caffau08hfth} on
the thorium and hafnium abundance in the Sun, and the Th/Hf chronometer pair
often serves as an example in the following sections.


\section{Formulae related to radioactive dating}

\subsection{Age uncertainties\label{s:errors}}

Any sample of radioactive nuclei decays according to the exponential law:
\begin{equation}
n_X(t)=n_{X0} \exp\left({ -t/\tau}\right)
\label{e:decayn}
\end{equation}
where $n_{X0}$ is the number of nuclei at $t$$=$$0$, and $n_X$ the number of
remaining nuclei at time $t$; $\tau$ is the \emph{mean lifetime}.
Explicitly, $t$ is given by 
\begin{equation}
t=\tau\ln{n_{X0}\over n_X} \, .
\label{e:tdec}
\end{equation}
Knowing the initial and present number of nuclei, one can obtain the time
interval that has elapsed since $t$$=$$0$.  In practice, to use the decay
processes to determine an age, one measures the number of nuclei of the
unstable species ($n_X$) relative to the number of nuclei of a stable element
which is related to $n_{X0}$. This is, e.g., the case for the Th/Hf pair.  Let
us call $n_R$ the number of nuclei of the stable reference species related to
$n_{X0}$. Let us further assume that the two nuclei have a common
nucleosynthetic origin and that the relation between the initial number of
nuclei $n_{X0}$ and $n_R$ is given by
\begin{equation}
n_{X0} = \beta n_R
\label{e:beta}
\end{equation}
where $\beta$ is the production ratio  between the respective numbers of nuclei. 
The production ratio  may be predicted by
theoretical considerations \citep[see][and references therein]{kratz07}. 
From Eq.~\eref{e:tdec} and Eq.~\eref{e:beta} one can write
\begin{equation}
t=\tau\ln{\beta n_R\over n_X} \, .
\label{e:tdec1}
\end{equation}

All astrophysically relevant unstable elements are trace species.
Let us therefore suppose that the lines corresponding to the transitions
we are interested in, are weak. In this case, the relation between 
equivalent width ($W$ hereafter) and abundance ($n$) is linear:
\begin{equation}
W=C\,n \hspace{5mm} \mathrm{or} \hspace{5mm} \log W = \log C + \log n,
\label{e:WC}
\end{equation}
$C$ being a constant defined by the line transition under consideration
(dependent on oscillator strength and excitation potential) and the thermal
structure of the stellar atmosphere under investigation.

Many lines -- including the \ion{Th}{ii} resonance line -- are blended with
  other species, and in practice the abundance is obtained by fitting a
  synthetic profile.  In this case the equivalent width is obtained
  theoretically from the best-fitting profile and the underlying abundances.
  The error on the equivalent width may then be obtained by propagating the
  error in Eq.~\eref{e:WC}.  One may also try to derive an equivalent width
  for the blended line by {\em subtracting} the theoretical equivalent width
  of the blending components from the total W of the blend. However, in
  general, this ignores saturation effects, and thus the ``corrected'' W
  differs from the true one. As an example we may consider the \ion{Th}{ii}
  resonance line in the Sun. Let us consider the theoretical computation of
  \citet{caffau08hfth} based on a 3D \cobold\ model: the W of the blend
  \ion{Th}{ii}+\ion{Co}{i}+\ion{V}{ii}, for an assumed abundance A(Th)=0.09 is
  0.74\,pm, the W of just the \ion{Co}{i}+\ion{V}{ii} blend is 0.31\,pm, while
  W(Th) = 0.46\,pm. In spite of the fact that all the lines involved are very
  weak, it is clear that the ``corrected'' W (0.43\,pm) differs from W(Th),
  due to saturation effects. Thus one should always use the equivalent width
  of the line, rather than the ``corrected'' W, although in many cases the
  difference is so small that it may be safely ignored. Alternatively, one may
  work with Eq.~\eref{e:sigtsiga} and use the obtained abundances
  directly. In this case one has to establish uncertainties by Monte-Carlo
  simulations.

Substituting Eqs.~\eref{e:beta} and \eref{e:WC} into Eq.~\eref{e:tdec} 
we obtain
\begin{eqnarray}
t &=& \tau\ln \left(\beta\,\frac{W_R}{W_X}\,\frac{C_X}{C_R}\right) \nonumber \\
  &=& \tau \left\{\ln \beta + \ln {W_R} - \ln {W_X} + 
  \ln \left(\frac{C_X}{C_R}\right) \right\} \, .
\label{e:tdec2}
\end{eqnarray}
The achievable precision in $t$ is hence limited by the uncertainty in the
knowledge of $\beta$, errors in the measurement of $W_X$ and $W_R$, as well as
by systematic errors in modeling the stellar atmosphere and line transitions,
i.e.\ errors of the constants $C_X$ and $C_R$.

From the standard formula for the propagation of uncorrelated errors
we get for the variance of $t$
\begin{eqnarray}
{\sigma(t)}=\tau \left\{\left({\sigma(\beta)\over \beta}\right)^2 + 
                        \left({\sigma(W_R)\over W_R}\right)^2 \right. &+&
                        \left({\sigma(W_X)\over W_X}\right)^2 \nonumber \\
&+& \left. \left({\sigma(C_X/C_R)\over C_X/C_R}\right)^2\right\}^{1/2}.
\label{e:sigt}
\end{eqnarray}
This equation is valid under the presumption that the errors of $\beta$,
$W_X$, $W_R$, and $(C_X/C_R)$ are uncorrelated. The errors of $C_X$ and $C_R$
themselves may often be more or less correlated.  
In fact, Th and Hf abundances are derived from singly ionized species
  which constitute the dominant ionization stage in the late-type stars of
  interest here. Moreover, the diagnostic lines of Th and Hf emerge from low
  lying energy levels which makes the Boltzmann factor and consequently the
  temperature sensitivity of the level populations modest.  The similar atomic
  properties of Th and Hf lead to similar formation regions in the
  photosphere.
Hence, all systematic errors due to uncertainties in gravity, effective
temperature and even the details of the thermal structure of the employed
model atmosphere should largely cancel out.  
Hf is the heaviest element prior to Th whose observable spectroscopic 
signatures are this favorable.
What is left is the systematic error related to the $gf$-values.

Taking the Sun and the Th/Hf pair as an example, we have ${\sigma(W_R) / W_R}
\approx 0.03/0.3 = 0.1$ (see \ion{Hf}{ii} $\lambda$~409.3\pun{nm} in Table~3
in \citet{caffau08hfth}), and ${\sigma(W_X) / W_X} \approx 0.03/0.4 = 0.075$ (see
Table~4 in \citealt{caffau08hfth}).  Since $^{232}$Th has a mean lifetime of
$\tau$$=$$20.3$~Gyr, we obtain from Eq.~\eref{e:sigt} under the most
optimistic conditions, assuming $\sigma(\beta)=0$ and $\sigma(C_X/C_R)=0$, an
error in the age determination of the order of ${\sigma(t)}$$\approx$$2.5$~Gyr.

Even this rather limited precision is probably beyond the present
state-of-the-art in stellar spectroscopy. However it is not implausible to
reach a precision of the order of 2\pun{Gyr} in the future.  The uncertainties
due to the S/N in the spectrum may be considerably reduced by the next
generation of ELTs, and uncertainties in the blending
lines can be reduced through laboratory measurements and/or theoretical
calculations.  The precision currently attainable on the dating of Globular
Clusters, using Main Sequence fitting, is a factor of two better (about
1\pun{Gyr}), however it may not be applied to field stars.  For field stars
which are evolved off the Main Sequence, if the distance is known, an age can
be derived through comparison with theoretical isochrones. However, distances
are difficult to measure and a small error in the distance propagates to a
great error in age.  In principle, radioactive dating has the potential to
provide ages of individual field stars, even if their distances are not
precisely known.

Unfortunately, in the literature equivalent widths and their uncertainties are
usually not given so that a direct application of the previous formulae is
not possible. In Sect.~\ref{s:decay} we will see that they are nevertheless
useful to derive a statement of the optimal decay time of a species for
measuring a particular time interval. For now, we proceed by listing the
previous relations in terms of the logarithmic abundances $A(X)=\log
\epsilon_X \equiv \log n_X/n_H + 12$.  Equation~\eref{e:tdec1} for the age
becomes
\begin{equation}
t=\tau\ln\!10 \left\{\log\beta + A(R) - A(X)\right\} , 
\label{e:tdec1a}
\end{equation}
with the corresponding uncertainty
\begin{eqnarray}
{\sigma (t)}&=&\tau\ln\!10 \left\{\rule{0mm}{1em} \sigma^2\!\left[\log\beta\right]+
\sigma^2\!\left[A(X)\right]+\sigma^2\!\left[A(R)\right]\right.\nonumber\\
\mbox{} &-& \left.2\,\sigma\!\left[A(X)\right]\,\sigma\!\left[A(R)\right]\,\cov{A(X)}{A(R)}\rule{0mm}{1em}\right\}^{1/2}\,.
\label{e:sigtsiga}
\end{eqnarray}  
$\cov{.}{.}$ denotes the linear correlation coefficient. In deriving
Eq.~\eref{e:sigtsiga}, it was assumed that $\beta$ is uncorrelated with
A(X) and A(R). Relation~\eref{e:sigtsiga} is consistent with
Eq.~\eref{e:sigt} as long as the statistical bias due to the change from
linear to logarithmic quantities is small, i.e. that for any involved
quantity~$q$
\begin{equation}
{\sigma(q)\over q} \approx \sigma(\ln\,q) = \ln\!10\,\,\sigma(\log\,q) .
\end{equation}  
\label{e:bias}
Equation~\eref{e:sigtsiga} now explicitly contains the correlation among the
measurements of the abundances of the two involved species. If abundance
errors are positively correlated -- as discussed previously and the Th/Hf pair
serving as example -- this can lead
to a substantial reduction of the overall error with respect to the
uncorrelated case. However, care has to be exercised when applying the
formulae to literature data (as we intend to do). Quoted abundance errors are
often purely statistical, uncorrelated uncertainties while correlated
uncertainties related to atmospheric models are specified separately by
providing sensitivities against changes of the stellar parameters.

\subsection{Optimal decay times\label{s:decay}}

\begin{figure}
\resizebox{8.8cm}{!}{\includegraphics[clip=true,angle=0]{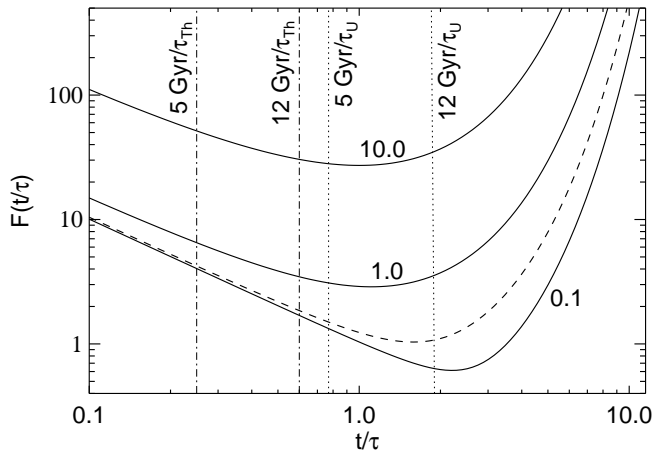}}
\caption{The function $F\left({t\over\tau};p\right)$ defined in
  Eq.~\eref{e:Faux} plotted as a function of ${t\over\tau}$ for different
  parameters~$p$ (solid curves). The dashed line shows the case
  of the Th/Hf chronometer pair under solar conditions (\ion{Th}{ii}
  $\lambda$~401.8\pun{nm} and \ion{Hf}{ii} $\lambda$~409.3\pun{nm} line pair,
  $p$$=$$0.26$).  The the dotted and dash-dotted vertical lines mark two characteristic ages,
  5 and 12\pun{Gyr}, taking the Th (20.3\pun{Gyr}) and U (6.5\pun{Gyr}) mean
  lifetimes as reference.}
\label{sigmat}
\end{figure}

We now discuss how the achievable precision depends on the time interval
$t$ we want to measure and the decay time of the radioactive species, $\tau$. 
From Eq.~\eref{e:sigt} or (\ref{e:sigtsiga}) it might appear that a radioactive 
species with short
mean lifetime $\tau$ provides the highest precision on $t$. However, this 
reasoning ignores the fact that a rapid decay ($\tau\ll t$) makes it more
difficult to measure $W_X$ precisely. To make this connection manifest
we eliminate $W_X$ and rewrite Eq.~\eref{e:sigt} as
\begin{eqnarray}
{\sigma(t)\over t}&=&{\tau\over t}\left\{\left({\sigma(W_R)\over W_R}\right)^2
\left(1 + \left({C_R\over \beta\,C_X}\right)^2\,\exp ({2t/\tau})\right) \right.
\nonumber \\ 
&+& \left. \left({\sigma(\beta)\over \beta}\right)^2 + 
 \left({\sigma(C_X/C_R)\over C_X/C_R}\right)^2\right\}^{1/2}\, ,
\label{e:sigt2}
\end{eqnarray}
where we have used the relations
\begin{equation}
{W_R\over W_X} = {C_R\over \beta C_X}\,\exp ({t/\tau})
\label{e:WxWn}
\end{equation}
and
\begin{equation}
\sigma(W_R) = \sigma(W_X)\, .
\label{e:cayrel}
\end{equation}
Eq.~\eref{e:WxWn} is obtained by combining Eqs.~\eref{e:decayn}, 
\eref{e:beta}, and \eref{e:WC}, while Eq.~\eref{e:cayrel} is 
a consequence of Cayrel's formula \citep{cayrel88}
which, for a spectrum that is photon noise dominated, states that the
uncertainty of the W measurement~$\sigma(W)$ depends on the FWHM of the
line, the pixel size, and the S/N of the spectrum. For spectra of moderate
resolution (R$\le 80000$), the FWHM is essentially the same for all lines
since they are not fully resolved and the FWHM is simply set by the
spectrograph's resolution. This is also true for higher resolution spectra if
the atoms under consideration do not have a very large difference in mass, and
therefore thermal broadening, or if the dominant broadening is non-thermal. In
the case of the Th/Hf pair, the mass difference is not large, and in late-type
stars the macro-turbulent plus rotational broadening is often of the same
order of magnitude or greater than the thermal broadening.  We may therefore
assume that $\sigma(W)$ is the same for any line in a given spectrum.
Assuming again $\sigma(\beta)$$=$$0$, and $\sigma(C_X/C_R)$$=$$0$,
and noting that $\sigma(W_R)/W_R$ is time independent,
the relative error of $t$ becomes
\begin{eqnarray}
{\sigma(t)\over t} &=& {\sigma(W_R)\over W_R}\,{\tau\over t}\,
\left(1+\left({C_R\over\beta C_X}\right)^2\exp
\left({2t/\tau}\right)\right)^{1/2} \nonumber \\
&\equiv&  {\sigma(W_R)\over W_R}\, F\left({t\over\tau};p\right)\,\,\,
\mathrm{with}\,\,\,p\equiv{C_R\over\beta C_X}\, .
\label{e:Faux}
\end{eqnarray}
For brevity, we introduced an auxiliary function~$F$ of argument
${t\over\tau}$ and parameter~$p$ in Eq.~\eref{e:Faux}.  The
smaller $F\left({t\over\tau}\right)$ at given $p$, the higher the precision of
the age determination.  The best (relative) accuracy is obtained for the time
where the function $F$ shows a minimum.
As shown in Fig.\,\ref{sigmat}, the function has a minimum near $t\approx
\tau$.  Left of the minimum ($t < \tau$), $F$ decreases slowly with time,
while it increases more steeply beyond the minimum ($t > \tau$).  We can
conclude that an optimal relative precision for radioactive dating is obtained
when one measures a time comparable to the mean lifetime of the unstable
element -- a perhaps intuitively expected result, put on a quantitative
footing here. 
Of course, our analysis is idealized by assuming a single, isolated line
  for both the stable and decaying species. In practice, often several lines
  are considered and line strengths are measured via spectral synthesis in
  order to account for line blending. However, we think that the overall
  picture in this more complex situation does not change qualitatively.

We finally consider radioactive dating when 
measuring two lines from two transitions of two unstable elements
which we call $X$ and $Y$.
There are two decay laws to be considered for the two elements:
\begin{equation}
n_X=n_{X0} \exp\left({-t/\tau_X}\right)~~~~{\rm and}~~~~
n_Y=n_{Y0} \exp\left({-t/\tau_Y}\right).
\end{equation}
We suppose that $\tau_X < \tau_Y$, and that the two elements are, 
at least partially, formed in the same nucleosynthetic
process, so that the abundances of nuclei of the two elements at the 
initial instant of time are related by
$n_{X0} = \beta n_{Y0}$.
In this way, we obtain
\begin{equation}
n_X=\beta n_Y\exp{\left(-t/\tau\right)}\, ,
\end{equation}
once $\tau$ is defined as
\begin{equation}
{1\over \tau}={1\over \tau_X} - {1\over \tau_Y}\, .
\label{e:taueff}
\end{equation}
We then find the same result for $\sigma(t)/t$ as obtained above
in Eq.~\eref{e:Faux}, except that  $\tau$ now depends on
the mean lifetimes of the two elements considered according to
Eq.~\eref{e:taueff}, and the reference element~$R$ is identified with the
longer lived species~$Y$ at its present abundance.

In practice, the only two radioactive nuclei which have been used to date
stars are $^{232}$Th and $^{238}$U with mean lifetimes of 20.3\pun{Gyr} and
6.5\pun{Gyr}, respectively.  From Eq.~\eref{e:taueff} the effective $\tau$ for
the Th/U pair amounts to 9.6\pun{Gyr}. From Fig.~\ref{sigmat} it is clear that
the Th/U pair as well as U/R pairs (R for any stable $r$-process element) are
well suited for measuring great ages around 12\pun{Gyr}, with Th/R being
competitive depending on the actual value of the $p$ parameter. Towards ages
of a few Gyr the sensitivity of U/R pairs is about 2 to 3 times higher than
the one of Th/R pairs, again depending on $p$. We repeat that only the decay
times enter the considerations here; it is usually more difficult
to measure U than Th abundances so that the advantage of the shorter life time
of U is offset. Finally, we ask whether there are other suitable naturally
occurring radioactive elements with lifetimes even more favorable for measuring
ages in the range of a few Gyr. The answer is `no': the lifetimes are either
much too short or much too long -- besides further complications like
exceedingly low abundances of the respective isotopes or the availability of
suitable spectral lines.


\section{Ages of metal-poor stars}

As an application of the formulae we derived for the errors on age, we
consider metal-poor stars for which the thorium abundance has been measured.
For all these stars the abundance of at least one other $r$-process element is
available (see Table~\ref{starabbo}).  Using various production ratios~$\beta$
available in the literature (see Table~\ref{pr}), we computed the ages of
these stars using Eq.~\eref{e:tdec1a} and their uncertainties with
Eq.~\eref{e:sigtsiga}. Correlations were neglected; correlations due to
uncertainties in stellar parameters are usually not included in the error
budget, and correlations emerging from the data reduction (e.g. background
reduction) not specified by the authors. Our results are summarized in
Table~\ref{starage}.  For star HE~1523-0901 the abundances of the single
elements were not available, but in \citet{frebel07} abundance ratios are
reported, and measurement errors on the individual abundances which is
sufficient for our purposes. By using different production ratios we intended
to separate the systematic uncertainties related to nucleosynthesis and
chemical evolution from the spectroscopic uncertainties as such. Since the
production ratios are treated separately here their related uncertainty in the
evaluation of $\sigma(t)$ with Eq.~\eref{e:sigtsiga} is set to zero,
$\sigma^2[\log\beta]=0$.


\begin{longtable}{llrrrrr}
  \caption{Ages and age uncertainties~$\sigma_{\rm Age}$ of metal-poor stars
    and the Sun obtained for different chronometer pairs~X/Y.\label{starage}}\\
\hline\noalign{\smallskip}
\hline\noalign{\smallskip}
X/Y & Object & \multicolumn{4}{c}{Age$^\mathrm{a}$} & $\sigma_{\rm Age}$\\
    &        & \multicolumn{4}{c}{[Gyr]} & [Gyr]\\
\hline\noalign{\smallskip}
\endfirsthead
\caption{continued.}\\
\hline\noalign{\smallskip}
\hline\noalign{\smallskip}
X/Y & Object & \multicolumn{4}{c}{Age$^\mathrm{a}$}   & $\sigma_{\rm Age}$\\
    &        & \multicolumn{4}{c}{[Gyr]} & [Gyr]\\
\hline\noalign{\smallskip}
\endhead
\hline\noalign{\smallskip}
\endfoot
\endlastfoot
Th/Eu & HD 115444      & 11.7  & 10.4  & 12.3  & 14.6 &  6.9\\
Th/Os & HD 115444      & 30.1  & 24.3  &        &       &  7.7\\
Th/Eu & CS 22892-052$^\mathrm{b}$ & 14.8  & 13.4  & 15.4  & 17.6 &  6.4\\
Th/Hf & CS 22892-052   & 27.7  &        &        &       &  7.4\\
Th/Os & CS 22892-052   & 19.6  & 14.8  &        &       & 18.5\\
Th/Eu & CS 22892-052   & 15.4  & 13.1  & 15.0  & 17.3 &  5.3\\
Th/Os & CS 22892-052   & 24.4  & 18.6  &        &       &  5.1\\
Th/Ir & CS 22892-052   & 25.8  & 19.1  & 25.2  &       & 12.9\\
Th/Eu & CS 22892-052   & 12.1  & 11.7  & 13.7  & 15.9 &  4.8\\
Th/Hf & CS 22892-052   & 24.6  &        &        &       &  6.1\\
Th/Os & CS 22892-052   & 26.0  & 20.3  &        &       &  7.7\\
Th/Ir & CS 22892-052   & 19.0  & 13.3  & 19.4  &       &  6.1\\
Th/Eu & BD +17 3248    &  7.6  &  6.3  &  8.3  & 10.5 &  5.3\\
Th/Os & BD +17 3248    & 27.7  & 22.0  &        &       &  6.1\\
Th/Ir & BD +17 3248    & 20.7  & 14.0  & 19.1  &       &  7.7\\
Th/Eu & HD 221170      & 11.7  & 10.4  & 12.3  & 14.6 &  4.2\\
Th/Hf & HD 221170      & 21.9  &        &        &       &  2.9\\
Th/Os & HD 221170      & 27.1  & 22.3  &        &       &  5.3\\
Th/Ir & HD 221170      & 20.7  & 14.0  & 19.1  &       &  6.1\\
Th/Eu & CS 31082-001   &  0.9  & -1.5  &  0.5  &  2.7 &  2.9\\
Th/Hf & CS 31082-001   & 15.2  &        &        &       &  8.0\\
Th/Os & CS 31082-001   & 17.9  & 12.1  &        &       &  5.9\\
Th/Ir & CS 31082-001   &  6.5  & -0.2  &  5.9  &       &  4.1\\
 U/Th & CS 31082-001   & 16.6  & 13.3  & 14.6  &       &  5.9\\
Th/Eu & HE~1523-0901   & 11.4  &  9.5  & 11.0  & 13.2 &  3.1\\
Th/Os & HE~1523-0901   & 16.9  & 10.1  &        &       &  3.1\\
Th/Ir & HE~1523-0901   & 18.0  & 12.3  & 17.4  &       &  3.1\\
 U/Th & HE~1523-0901   & 15.0  &        &        &       &  2.6\\
Th/Eu & M51 K341       & 11.0  &  9.7  & 12.6  & 13.9 &  6.9\\
Th/Eu & M51 K462       & 14.1  & 12.7  & 14.7  & 16.9 &  6.9\\
Th/Eu & M51 K583       &  6.6  &  4.3  &  7.2  &  8.5 &  6.9\\
Th/Hf & M4  L1411     & 32.5  &       &       &      &  3.4\\
Th/Hf & M4  L1411     & 32.5  &       &       &      &  3.4\\
Th/Hf & M4  L1501     & 25.5  &       &       &      &  3.4\\
Th/Hf & M4  L1514     & 37.2  &       &       &      &  3.4\\
Th/Hf & M4  L2406     & 32.5  &       &       &      &  3.4\\
Th/Hf & M4  L2617     & 23.2  &       &       &      &  3.4\\
Th/Hf & M4  L3209     & 30.2  &       &       &      &  3.4\\
Th/Hf & M4  L3413     & 23.2  &       &       &      &  3.4\\
Th/Hf & M4  L4511     & 27.8  &       &       &      &  3.4\\
Th/Hf & M4  L4611     & 37.2  &       &       &      &  3.4\\
Th/Hf & M5  IV-81     & 18.5  &       &       &      &  3.4\\
Th/Hf & M5  IV-82     & 18.5  &       &       &      &  3.4\\
Th/Eu & HD 108577  &  7.8  &  6.2  &  8.4  & 10.1  &  3.4\\
Th/Eu & HD 115444  &  9.6  &  8.1  & 10.3  & 11.9  &  3.6\\
Th/Eu & HD 186478  & 16.6  & 15.1  & 17.3  & 18.9  &  4.3\\
Th/Eu & BD +82856  &  7.3  &  5.8  &  8.0  &  9.6  &  5.5\\
Th/Eu & M92 VII-18 &  6.8  &  5.3  &  7.5  &  9.1  &  5.3\\
Th/Eu & Sun &  3.3 &  1.7 &  3.9 &  5.6 & 2.0\\ 
Th/Hf & Sun & 22.3 &      &      &      & 2.3\\
Th/Os & Sun & 14.1 &  8.6 &      &      & 4.9\\
Th/Ir & Sun & 10.1 &  4.0 &  9.6 &      & 4.9\\
\noalign{\smallskip}
\hline
\end{longtable}
{\scriptsize\noindent
{$^\mathrm{a}$\,Ages and age uncertainties are obtained from
    formulae~\eref{e:tdec1a} and \eref{e:sigtsiga} using the abundances listed in Table~\ref{starabbo}. 
    The four ``Age'' columns give ages based on production rations~$\beta$ of \citet{kratz07}, \citet{sneden03}, \citet{schatz02}, and
    \citet{cowan02} (in that order) according to Table~\ref{pr}. The Solar ages are
    calculated applying $r$-process fractions as given in
    Table~\ref{rfraction}.}\\
{$^\mathrm{b}$\,Three independent measurements were obtained for this star which is
the reason why certain chronometer pair/object combinations are listed multiple
times.}}

\twocolumn

We like to point out that we applied the spectroscopy-related uncertainties of
the abundances as given by the authors. There is no general consensus about
the appropriate way to estimate the error on abundances, and the approach
varies from author to author. For those elements which have several measurable
lines the error can be estimated by looking at the RMS line-to-line abundance
scatter. One further might divide this error by $\sqrt{N}$ where ${N}$ is the
number of lines, provided the errors on each line are independent.  However,
this assumption is often questionable, since in most cases the lines are
measured from the same spectrum. Any error which affects the whole spectrum
(e.g. an error in the background subtraction) will affect all the lines in the
same (or at least in a similar) way. We thus argue that it is more realistic
and conservative not to invoke this factor $1/\sqrt{N}$.

\begin{figure}
\includegraphics[width=8cm]{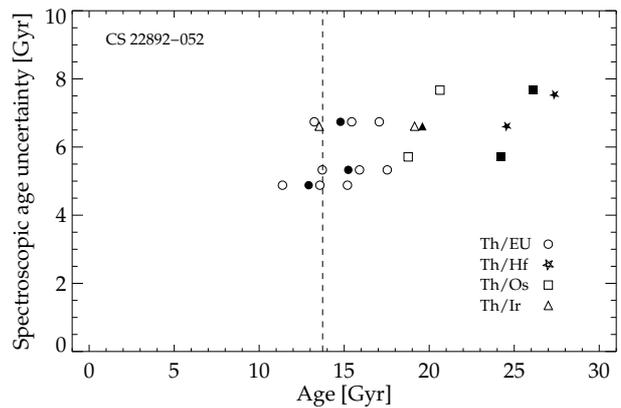}
\caption{Ages and spectroscopic age uncertainties for star CS~22892-052
  determined from various  chronometer pairs (symbols) assuming up to four different
  production ratios. Filled symbols refer to the
  production rations of \citet{kratz07}. The dashed line indicates the age of
  the universe. \citet{sneden03} give a radiochronometric age estimate of $14.2\pm
  3$\pun{Gyr} for this star.}
\label{snedenstar}
\end{figure}

\begin{figure}
\includegraphics[width=8cm]{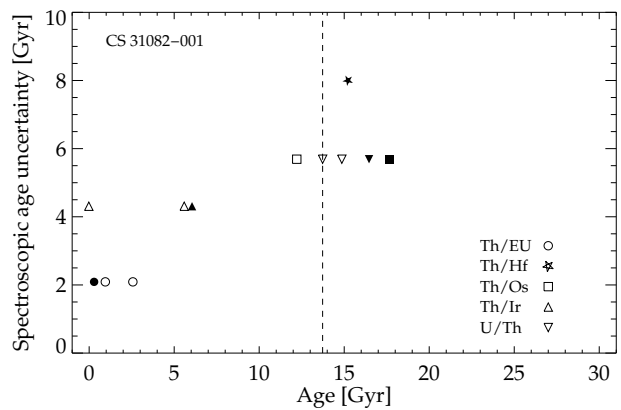}
\caption{Same as Fig.~\ref{snedenstar} for star CS 31082-001. In addition, for
  this star U/Th ages are available. \citet{hill02} give a  radiochronometric age estimate
  of $14.0\pm 2.4$\pun{Gyr} for this stars.}
\label{hillstar}
\end{figure}

The ages of the individual metal-poor stars in Table~\ref{starage} show a great
spread when considering different $r$-production ratios, $\beta$, and chronometer 
pairs. Also the age uncertainties stemming from the spectroscopically determined
abundances, $\sigma_{\rm Age}$, are substantial. We note that our 
spectroscopy-related uncertainties
are commonly greater than the estimates given by the authors themselves.
Figures~\ref{snedenstar} and~\ref{hillstar} provide a graphical representation
of the situation for the stars CS~22892-052 and CS~31082-001, respectively. We
picked these objects since spectroscopic measurements for a greater number of
chronometer pairs exist. CS~31082-001 is of particular interest since it is
strongly $r$-process element enhanced allowing the measurement of the Th
abundance exceptionally from several lines making it more reliable.
For CS~22892-052 we left out the age estimates with the highest uncertainties,
stemming from Th/Os and Th/Ir, since they were later improved substantially. 
Both figures
illustrate that the dispersion due to different production ratios is often
greater than the dispersion due to abundance uncertainties. This is even the
case if one restricts the comparison to production ratios originating from the
same author. As already pointed out in the original paper on the abundance
analysis of CS~31082-001 by \citet{hill02} the age derived from the Th/Eu pair
is obviously unrealistic despite rather accurate abundance measurements. The
same holds to lesser extend for the Th/Ir pair. This points to shortcomings in
our understanding of the $r$-process production ratios and/or the chemical
history of the star. For CS 22892-052 all the reference elements except Eu
lead to exceedingly great ages. One might speculate that the
Th/Hf age might be enlarged by an $s$-process contribution to the Hf abundance:
although the star has a very low metalicity, it is also carbon enhanced, so
that one cannot rule out the possibility that it may have a (so far
undetected) companion which went through the AGB phase, polluting the
photosphere of the primary. However, even this hypothesis cannot be invoked to
explain the great Th/Os and Th/Ir ages.  Os and Ir are in fact supposed to
show only minor $s$-process contributions \citep{arlandini99}. Again, we must
conclude that large systematic uncertainties still exist in the theoretical
calculations of production ratios~$\beta$.

At this point one may ask what accuracy has been reached in
  spectroscopically measured abundance ratios necessary for radioactive dating
  -- in particular for metal-poor stars --, and what are the perspectives for
  improving these accuracies in the future. Both questions are not easy to
  answer. Uncertainties given by individual authors are often on the
  optimistic side. We think that a comparison among different authors provides
  a more robust measure of the actually present -- as opposed to published --
  uncertainties. Such a comparison might still be biased towards too small
  error bars since different authors often draw from the same sources for
  oscillator strengths, or may use the same observational material.  Hence,
  the analyses are not independent. \citet{sneden09} performed such a
  comparison of the abundances which they obtained for the metal-poor
  $r$-process enriched halo giant CS~31082--001 with studies of \citet{hill02}
  and \citet{honda04}. We repeat this comparison here for the obtained
  \textit{abundance ratios}. In the comparison Honda--Sneden the list of
  common elements is La, Ce, Pr, Nd, Sm, Eu, Gd, Tb, Dy, Er, Tm, Yb, in
  Hill--Sneden La, Ce, Pr, Nd, Sm, Eu, Gd, Tb, Dy, Er, Tm, Hf. Abundance
  ratios among these elements can be taken as a rather good model for the Th/Hf
  pair since their lines emerge from the first, dominant ionization stage, and
  from levels of low excitation potential \citep{lawler09,sneden09} -- similar
  to the situation for the Th/Hf pair.  Taking the abundance ratios from the
  individual studies ensures that correlation effects reducing the errors on
  abundance ratios are fully taken into account. For the rare earth elements
  often many lines are available for the determination of their abundance.
  This luxury is not afforded by Th or Hf so that we still have a likely bias
  towards too high accuracy when taking the rare earths as example.

From the lists of elements in common we formed all possible abundance ratios
(except for an element with itself), and calculated the difference of
the logarithmic ratios between the various studies. For the pair Hill--Sneden
we find a dispersion among the differences of the logarithmic ratios of
0.14\,dex, for Honda--Sneden 0.12\,dex. In the case of Hill--Sneden, in
particular the discrepant Hf abundances have to be noted. In the case of
Honda--Sneden, a substantial part of the dispersion is driven by the elements
Tm and Tb; leaving them out reduces the dispersion to 0.05\,dex; however, this
is not suggested by, e.g., excessive errors on their abundance as stated by
the authors. We emphasize that at least in part the various groups use the
same oscillator strengths. From this exercise we conclude that attributing an
uncertainty of about 0.1\,dex to the Th/Hf abundance ratio is perhaps
conservative but not grossly over-estimating the actual value.

How to improve the accuracy? \citet{sneden09} report a line-to-line scatter of
typically 0.07\,dex in a given star when determining abundances of a rare
earth element and applying the high-quality oscillator strength of
\citet[][and references therein]{lawler09}. A sub-optimal choice of the model
atmosphere introduces some scatter, but certainly errors in the oscillator
strength are an important contributor. Since two abundances are involved in
the determination of a ratio, an uncertainty of about 0.1\,dex would result
when an individual abundance is given to 0.07\,dex accuracy. It is often
argued that by averaging over abundances of $N$ individual lines one improves
the obtained accuracy by a factor~$\sqrt{N}$. However, this would only be the
case if the scatter among the oscillator strength is of random nature which is
far from obvious. So, even more accurate oscillator strengths of the involved
line transitions are desirable. This also relates to lines of blending species
which one has to deal with. At this point an appropriate theoretical
description of the line formation process also comes into play which should
account for departures from local thermodynamic equilibrium and gas-dynamical
effects. All this is alleviated by having high signal-to-noise and high
resolution spectra at hand which ELTs will be able to provide. One often has
to work with spectra where a line profile is sampled with only very few
points. Better sampling usually permits more accurate measurements. We are
convinced that we are not stuck at the present level of accuracy to which
abundance ratios used in radioactive dating can be determined but not
insignificant effort is necessary to make substantial progress.

%

\section{The Solar age}

As another application of the dating formulae we tried to deduce the solar
age. We start by taking the Th/Hf pair as an example since we worked recently
ourselves on this chronometer pair \citep{caffau08hfth} having all relevant
spectroscopic information at hand. In the case of the 401.9\pun{nm}
\ion{Th}{ii} / 409.3\pun{nm} \ion{Hf}{ii} line pair, our favorite spectral
synthesis code gives ${C_R / C_X}=0.13$.  For the Sun one has to consider two
complications which are less severe for metal-poor stars:
\begin{enumerate}
\item
the Th and Hf observed in the solar system are not
the result of a single $r$-process event, but rather
of the superposition of many events;
\item while Th is produced only through the $r$-process,
Hf is in part produced by the $s$-process.
\end{enumerate}
To take into account the first point one should use an appropriate model of
Galactic chemical evolution \citep[see, e.g.,][]{delpeloso2}.  We ignore this
aspect and assume a single $r$-process event, to see if at least a reasonable
estimate of the solar age can be derived.  The second point instead can easily
be taken into account by using the solar $r$-process fraction of Hf nuclei.
This has been estimated by \citet{lawler07} to be 56.6\pun{\%} of the total Hf
nuclei. For the production ratio we assume $\beta=0.862$, from Table 2 of
\citet{kratz07}. The curve $F\left({t\over\tau};p=0.26\right)$ related to this
case is depicted as a dashed line in Fig.~\ref{sigmat}.  For ages 5 and
12\pun{Gyr} (i.e. for times corresponding to the age of the Sun and to the age
of Galactic Globular Clusters, respectively), we find ${\sigma(t)/t} = 4.3
\times {\sigma(W_R)/W_R}$ and $1.9 \times {\sigma(W_R)/W_R}$, respectively.
We give these numbers mainly for illustrating the application of the formulae
involving the equivalent widths. In the solar case they are not strictly
applicable since when deriving Eq.~\eref{e:Faux} we assumed that uncertainties
in the observed spectra are related to the photon noise. This is certainly not
the case for Solar spectra.  The uncertainties are related to the systematics,
in particular blending of the thorium line.

Turning now to spectroscopically measured abundances, Table~\ref{starage}
lists the obtained ages and their uncertainties, again for different
chronometer pairs and assumptions about production ratios and $r$-process
fractions (see Tables~\ref{starabbo}, \ref{pr}, \ref{rfraction}). The Th/Hf
age, 22.3\pun{Gyr}, comes out to be greater than the age of the universe,
13.73\pun{Gyr}, as estimated from the fluctuations of the Cosmic Microwave
Background \citep{spergel07}.  
By using the A(Hf)=0.88 value suggested by \citet{lawler07}, only 0.01\,dex
higher than the abundance we use, the age comes out even a bit greater by 0.5\,Gyr.
Even if the solar abundance of Hf and Th were
produced in a number of distinct supernova events at different times, the
resulting \emph{mean} age as computed from Eq.~\eref{e:tdec1a} cannot be
greater than the time elapsed since the earliest supernova event, which cannot
have occurred earlier than 13.73\pun{Gyr} ago.  We have ignored the
destruction of Th by photons in stellar interiors. According to
\citet{delpeloso2} the matter cycled through stars may have a reduction of
10--20\% in Th. However, even when increasing the Solar Th abundance by 20\%, 
one still derives an age of 18.2\pun{Gyr} from Eq.~\eref{e:tdec}. 
We have to conclude that either the assumed value of
$\beta$, or the $r$-process fraction of Hf is incorrect.  Among the other
chronometer pairs Th/Eu provides the closest match to the actual Solar age,
while in general -- like for Th/Hf -- there it the tendency to grossly
overestimate the age. 
The photospheric solar abundance of europium provided by \citet{lawler07}
is exactly the same we are using.
Obviously, the complex chemical history of the Sun together
with uncertainties in the production ratios make it very difficult to obtain a
reliable Solar age by radioactive dating. 

In this context it is perhaps interesting to note that \citet{sneden09} find that
  the solar photospheric Hf abundance is deviating from the meteoritic value
  in contrast to all its preceeding (in atomic number) rare earth
  elements. Similar, the solar $r$-process abundance pattern corresponds to
  the one of $r$-process enriched iron-poor halo stars except for Hf which
  appears relatively depleted in the Sun. Of course, a higher $r$-process
  abundance of Hf in the Sun would even lead to a greater solar age aggravating
  the problem in the radioactive dating.

\section{Conclusions and remarks}

The accuracy of radioactive cosmo-chronometry is not going to be extremely
high on a single star.  A realistic estimate of what future instrumentation
will be able to achieve does likely not allow one to hope for a better than
2\pun{Gyr} precision, taking into account {\em observational} uncertainties
only, leaving aside the systematics associated to production rates and
chemical enrichment models. Although high-mass n-capture elements are
currently believed to be produced in a fairly well understood manner by the
so-called ``main'' $r$-process \citep[][]{farouqi08,kratz07}, the quantitative
determination of the production ratios $\beta$ between the stable and the
unstable elements still bear significant uncertainties.  


Since in photon noise dominated spectra the measurement error on the
equivalent width decreases linearly with the observational S/N ratio, the
attainable dating precision increases roughly linearly with the S/N ratio.
This constitutes a major practical challenge to the application of the method,
since many of the involved lines are in the blue part of the optical spectrum
(e.g. for Th, U, and Hf), and the natural target of the analysis are cool
stars.

If possible, the decaying element should be chosen such that its mean lifetime
$\tau$ is similar to the age one wants to measure.  This applies also to the
``effective'' $\tau$ of a chronometric pair constituted by two decaying
species. From this perspective, the pairs U/Th, Th/stable, and U/stable are
well suited to measure ages of old objects around 10\pun{Gyr} of age. No other
naturally occurring isotope has a mean lifetime providing a similarly close
match.

Potentially important systematic spectroscopy-related effects that we feel
deserve further study to fully exploit the potentials of radioactive dating
are effects related to departures from thermodynamic equilibrium and
gas-dynamics on the line formation of the chronometric species. For instance,
the application of a 3D radiation-hydrodynamical model atmosphere by
\citet{caffau08hfth} led to a downward revision of the photospheric Solar Th
abundance by 0.1\pun{dex}. This corresponds to a decrease of Th ages by
4.7\pun{Gyr}. 


Globular clusters provide fairly large samples of metal-poor, bright, coeval
and chemically homogeneous stars. Simultaneous measurement of abundances in
these stars allows us to reduce the observational, statistical uncertainties.
This could permit to derive $r$-process production ratios empirically by
calibration against the known ``photometric'' age. To our knowledge, this has
never been attempted so far.


\begin{acknowledgements}
  The authors H.-G.L., P.B, and L.S. acknowledge financial
  support from EU contract MEXT-CT-2004-014265 (CIFIST).
\end{acknowledgements}

\appendix

\section{Tables of elemental abundances, production ratios, and solar
  $r$-process fractions}

The following tables list the abundances, production ratios, and solar system
$r$-process fractions that we compiled from the literature for the purpose of
radioactive dating. In Table~\ref{starabbo}, exclusively spectroscopically
determined (as opposed to meteoritic) abundances are given for the Sun. In
Table~\ref{pr} of production ratios, the values for \citet{kratz07} refer to
the ones listed as ``Fe-Seed fit2'' in Table~2 of their paper. Kratz and
collaborators report three different estimates of the various production
ratios.

\begin{table*}
\caption{$r$-process elemental abundances in metal-poor stars and the Sun for
  radioactive dating.\label{starabbo}}
{\scriptsize
\begin{center}
\begin{tabular}{lllllllll}
\hline
\noalign{\smallskip}
 Object & Reference & [Fe/H] & A(Eu) & A(Hf) & A(Os) & A(Ir) & A(Th) & A(U)\\ 
\noalign{\smallskip}
\hline
\noalign{\smallskip}
HD 115444  &  \citet{cowan99} & --2.7 & $-1.5$  &  &  &   & $-2.1$  &    \\
HD 115444  &  \citet{westin00} & --2.9 &  $-1.63\pm 0.07$ &   & $-0.55\pm 0.11$ &  & $-2.23\pm 0.11$ & $<-2.6$\\
%
CS 22892-052 & \citet{cowan99} & --3.1 & $-0.9$ &  &  &   & $-1.6$ &   \\
CS 22892-052 & \citet{sneden00a}$^\mathrm{a}$
 & --3.1 & $-0.89\pm 0.12$ & $-0.90\pm 0.14$ & $-0.10\pm 0.38$ &  & $-1.55\pm 0.08$ & \\
CS 22892-052 & \citet{sneden00a} & --3.1 & $-0.93\pm 0.09$ &   & $-0.05\pm 0.10$ & $+0.00\pm 0.26$ &  $-1.60\pm 0.07$ & \\
CS 22892-052 & \citet{sneden03}  & --3.1 & $-0.95\pm 0.03$ & $-0.98\pm 0.10$ & $+0.02\pm 0.13$ & $-0.10\pm 0.10$ & $-1.57\pm 0.10$ & $<-2.3$\\
%
BD +17 3248 & \citet{cowan02} & --2.0 & $-0.67\pm 0.05$ &  & $+0.45\pm 0.10$ & $+0.30\pm 0.13$& $-1.18\pm 0.10$ & $-2.0:\pm 0.30$\\
%
HD 221170 & \citet{ivans06} & --2.18 & $-0.86\pm 0.07$ & $-0.94\pm 0.04$ & $+0.16\pm 0.10$ & $+0.02\pm 0.13$ & $-1.46\pm 0.05$ &  \\
%
CS 31082-001 & \citet{hill02} & --2.90 & $-0.63\pm 0.04$ & $-0.59\pm 0.17$ & $+0.43\pm 0.12$ & $+0.20\pm 0.09$ & $-0.98\pm 0.02$ & $-1.92\pm 0.11$\\
%
%
M51 K341 & \citet{sneden00b} & --2.32 & $-0.88\pm 0.09$ &    &    &   & $-1.47\pm 0.10$ & \\ 
M51 K462 & \citet{sneden00b} & --2.25 & $-0.61\pm 0.09$ &    &    &   & $-1.26\pm 0.10$ & \\ 
M51 K583 & \citet{sneden00b} & --2.34 & $-1.22\pm 0.09$ &    &    &   & $-1.7:\pm 0.10$ & \\ 
%
M4  L1411 & \citet{yong08}  & $-1.23$ & & $+0.08$ & & & $-0.68$ & \\
M4  L1501 & \citet{yong08}  & $-1.29$ & & $-0.02$ & & & $-0.63$ & \\
M4  L1514 & \citet{yong08}  & $-1.22$ & & $+0.03$ & & & $-0.83$ & \\
M4  L2406 & \citet{yong08}  & $-1.30$ & & $-0.12$ & & & $-0.88$ & \\
M4  L2617 & \citet{yong08}  & $-1.20$ & & $+0.08$ & & & $-0.48$ & \\
M4  L3209 & \citet{yong08}  & $-1.25$ & & $+0.03$ & & & $-0.68$ & \\
M4  L3413 & \citet{yong08}  & $-1.23$ & & $-0.02$ & & & $-0.58$ & \\
M4  L4511 & \citet{yong08}  & $-1.22$ & & $+0.03$ & & & $-0.63$ & \\
M4  L4611 & \citet{yong08}  & $-1.09$ & & $+0.18$ & & & $-0.68$ & \\
M5  IV-81 & \citet{yong08}  & $-1.28$ & & $-0.12$ & & & $-0.58$ & \\
M5  IV-82 & \citet{yong08}  & $-1.33$ & & $-0.22$ & & & $-0.68$ & \\
%
HD 108577 & \citet{johnson01}  & $-2.38$  &  $ -1.48\pm 0.02$ & & & &  $-1.99\pm 0.07$ & \\
HD 115444 & \citet{johnson01}  & $-3.15$  &  $ -1.81\pm 0.03$ & & & &  $-2.36\pm 0.07$ & \\
HD 186478 & \citet{johnson01}  & $-2.61$  &  $ -1.56\pm 0.06$ & & & &  $-2.26\pm 0.07$ & \\
BD +82856 & \citet{johnson01}  & $-2.12$  &  $ -1.16\pm 0.06$ & & & &  $-1.66\pm 0.10$ & \\
M92VII-18 & \citet{johnson01}  & $-2.29$  &  $ -1.48\pm 0.09$ & & & &  $-1.97\pm 0.07$ & \\
%
%
Sun  & \citet{grevesse98}    & +0.0 & & & $+1.45\pm 0.10$ & $+1.35\pm 0.10$\\
Sun  & \citet{mucciarelli}   & +0.0 & $+0.52\pm 0.03$ \\
Sun  & \citet{caffau08hfth}  & +0.0 & & $+0.87\pm 0.04$ & & & $+0.08\pm 0.03$\\
\noalign{\smallskip}
\hline
\end{tabular}
\end{center}
$^\mathrm{a}$\,Sneden et al. analysed two
different spectra for CS~22892-052 providing two independent
measurements. For this reason the object/reference combination is listed twice.
}
\end{table*}

\begin{table}[h!]
\caption{$r$-process production ratios.\label{pr}}
\begin{center}
\begin{tabular}{lllllll}
\hline
\noalign{\smallskip}
Reference & \multicolumn{5}{c}{Production ratio~$\beta$} \\
     & Th/Eu & Th/Hf & Th/Os & Th/Ir & U/Th\\
\noalign{\smallskip}
\hline
\noalign{\smallskip}
\citet{kratz07} & 0.453 & 0.862 & 0.093 & 0.089 & 0.638\\
\citet{sneden03}            & 0.420\\
\citet{schatz02}            & 0.468 &       & 0.071 & 0.066\\
\citet{cowan02}             & 0.507 &       &       & 0.087\\
\noalign{\smallskip}
\hline
\end{tabular}
\end{center}
\end{table}

\begin{table}
\caption{Solar $r$-process fractions~$f_\mathrm{r}$ for various elements~X.\label{rfraction}}
\begin{center}
\begin{tabular}{lll}
\hline
\noalign{\smallskip}
X & $f_\mathrm{r}$ & Reference\\
\noalign{\smallskip}
\hline
\noalign{\smallskip}
Eu & 0.942 & \citet{arlandini99}\\
Hf & 0.566 & \citet{lawler07}\\
Os & 0.92 & \citet{simmerer04}\\
Ir & 0.99 & \citet{simmerer04}\\
\noalign{\smallskip}
\hline
\end{tabular}
\end{center}
\end{table}

\bibliographystyle{aa}

\end{document}